\documentclass[runningheads, anonymous]{llncs}
\usepackage{amsmath,graphicx}
\usepackage{booktabs}
\usepackage{multirow}
\usepackage{makecell}
\usepackage{amssymb}
\usepackage{cite}
\usepackage{marvosym}
\begin{document}
%

\title{SPPNet: A Single-Point Prompt Network for Nuclei Image Segmentation}
%
%
%
\author{Qing Xu\inst{1,2} \and
Wenwei Kuang\inst{2} \and
Zeyu Zhang\inst{2} \and
Xueyao Bao\inst{2} \and
Haoran Chen\inst{2} \and
Wenting Duan\inst{1}
}
\authorrunning{Qing Xu et al.}
%
\institute{University of Lincoln, UK \and
University of Hong Kong, Hong Kong \\
\email{xq141839, kuangww, baogela, zeyu2022, rickchen@connect.hku.hk} 
\email{	wduan@lincoln.ac.uk}}

\titlerunning{A Single-Point Prompt Network for Nuclei Image Segmentation}
\maketitle              
\begin{abstract}
Image segmentation plays an essential role in nuclei image analysis. Recently, the segment anything model has made a significant breakthrough in such tasks. However, the current model exists two major issues for cell segmentation: (1) the image encoder of the segment anything model involves a large number of parameters. Retraining or even fine-tuning the model still requires expensive computational resources. (2) in point prompt mode, points are sampled from the center of the ground truth and more than one set of points is expected to achieve reliable performance, which is not efficient for practical applications. In this paper, a single-point prompt network is proposed for nuclei image segmentation, called SPPNet. We replace the original image encoder with a lightweight vision transformer. Also, an effective convolutional block is added in parallel to extract the low-level semantic information from the image and compensate for the performance degradation due to the small image encoder. We propose a new point-sampling method based on the Gaussian kernel. The proposed model is evaluated on the MoNuSeg-2018 dataset. The result demonstrated that SPPNet outperforms existing U-shape architectures and shows faster convergence in training. Compared to the segment anything model, SPPNet shows roughly 20 times faster inference, with 1/70 parameters and computational cost. Particularly, only one set of points is required in both the training and inference phases, which is more reasonable for clinical applications. The code for our work and more technical details can be found at https://github.com/xq141839/SPPNet.

\end{abstract}
\section{Introduction}

Automatic nuclei segmentation has received significant attention from pathologists and become an indispensable part of computer-aid diagnosis in the future \cite{wang2022medical}. In the last few years, deep neural networks displayed outstanding performance in various computer vision tasks. Specifically. U-Net, proposed by Ronneberger et al. \cite{ronneberger2015u}, has achieved great success in biomedical image segmentation. It adopts a convolutional encoder and decoder to extract image features and recover the original image shape respectively. A skip connection strategy is used to incorporate low-level semantic information with high-level semantic information in each layer. Zhou et al. \cite{zhou2019unet++} designed a nested U-Net (Unet++). It adds a series of nested and dense skip pathways between the encoder and decoder so that each layer can receive multi-scale semantic information. Schlemper et al. \cite{schlemper2019attention} developed an attention U-Net that uses a bottom-up attention gate to weight the concatenated feature map of the encoder and decoder. It highlights target regions and improves the sensitivity of the model. Instead of using a solely convolutional neural network (CNN) to encode image features, Chen et al. \cite{chen2021transunet} proposed TransUNet that combines CNN with vision transformer (ViT) \cite{dosovitskiy2020image} for image patch extraction and global context collection. With the increasing influence of transformer, a growing number of modern medical segmentation models embed ViT or even adopt fully-transformer architecture \cite{yan2022after, cao2022swin}. Recently, segment anything model (SAM) \cite{kirillov2023segment} was proposed as a foundation model for natural image segmentation. It consists of three modules: a large masked autoencoders (MAE) \cite{he2022masked} pre-trained ViT as image encoder. A prompt encoder is used to fetch tokens from box, point, or text inputs. A two-layer mask decoder leverages cross-attention for updating both the image embedding and prompt tokens. However, it can be difficult to retrain or fine-tune SAM when limited training sources are available. For nuclei image segmentation, although SAM can present remarkable performance when provided sufficient cell box or point prompts \cite{he2023accuracy}, it is not efficient for clinical applications. 

In this paper, we propose a single-point prompt network (SPPNet) for nuclei segmentation in microscopic images. In order to improve the model robustness in practical scenarios, a center neighbour selection algorithm is established using the combination of distance transform and Gaussian kernel. To reduce the parameters of image encoder, the MAE ViT of SAM is replaced with a Tiny-ViT \cite{wu2022tinyvit}. This small network may be not able to provide feature representations as complex as the large one. Therefore, we introduce a helpful shadow network, named low-level semantic information extractor (LLSIE). We evaluate the proposed on the small dataset: MoNuSeg-2018 \cite{kumar2017dataset}, and train the model with only one RTX2080Ti graphic card. The experimental result shows that SPPNet performs better than other state-of-the-art (SOTA) models and takes more than 20\% less training time. Compared to SAM, our proposed model requires much lower parameters, computational cost, and faster inference speed. Overall, SPPNet shows promising potential as a new cell segmentation method for pathological images in clinical settings.

\section{Method}

\subsection{Center Neighborhood Point Sampling}

\begin{figure}[!thbp]
  \centering
  \includegraphics[width=0.95\linewidth]{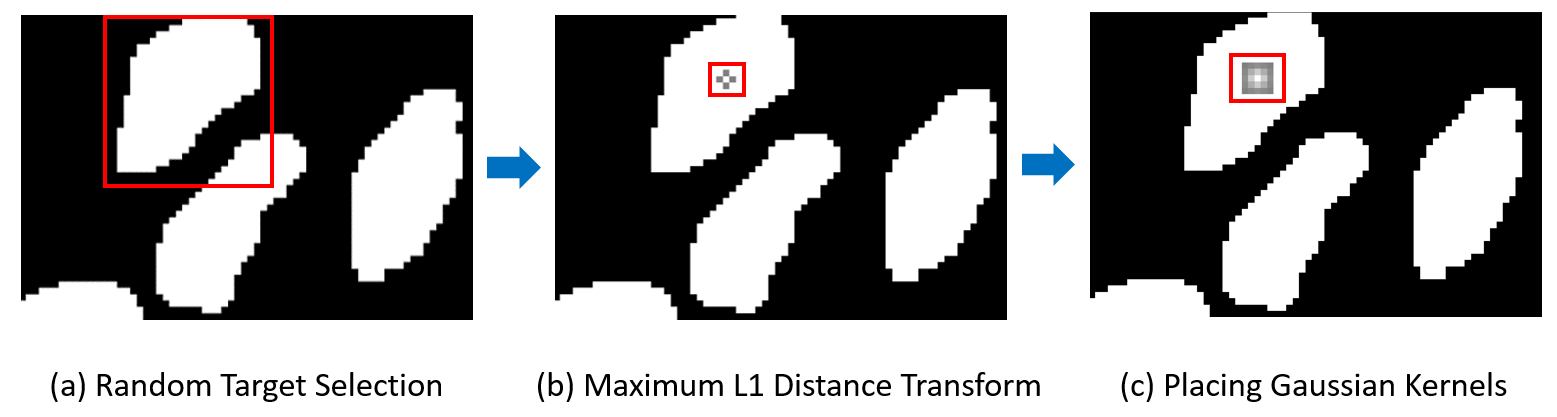}
  \caption{Algorithm of center neighborhood point selection.}
  \label{fig:f1}
\end{figure}

The SAM model supports box and point prompts as inputs to direct the network focus on target regions and provide precise segmentation masks. The former requires gathering all potential cell bounding boxes in images. The latter expects to acquire the center point of cells. Both methods are difficult to be implemented in practice as one whole-slide microscopic image may include thousands of nuclei and medical experts cannot guarantee to always label the center of cells. To address this issue, we design a center neighborhood point sampling algorithm, which is presented in Fig. \ref{fig:f1}.

Firstly, we randomly select one cell from the ground truth as the target region $\boldsymbol{x}$. Secondly, to find the center point $M$ of the nuclei, the L1 distance transform method is applied to this area, which can be represented as:
\begin{equation}
M(\boldsymbol{x})=\arg \max \left[\min _{\boldsymbol{y} \in \Omega}|\boldsymbol{x}-\boldsymbol{y}|\right]
\end{equation}
The transform calculates the minimum L1 distance between each point in $\boldsymbol{x}$ and the cell border $\Omega$. Then, we select the point with the maximum value in this set as the center of the cell. If more than one point satisfies the condition, we choose the first point in order. Thridly, a Gaussian filter is operated on this center to capture its neighborhood points. They can be defined as:
\begin{equation}
G_{\sigma}\left(m_{x}, m_{y}\right)=C \cdot e^{-\frac{m_{x}^{2}+m_{y}^{2}}{2 \sigma^{2}}}, m_{x}, m_{y} \in\left\{-K, \ldots, 0, \ldots, K\right\}
\end{equation}
Where $G_{\sigma}\left(m_{x}, m_{y}\right) \in \mathbb{R}^{\left(2 K+1\right) \times\left(2 K+1\right)}$ stands for a 2D Gaussian kernel and $\sum_{m_x=-K}^K\sum_{m_y=-K}^K G_{\sigma}\left(m_{x}, m_{y}\right) =1$. $\sigma^2$ is the isotropic covariance, the kernel size can be defined as $(2K + 1) \times (2K + 1)$ and $C$ is a constant. Finally, we randomly pick one from these neighbors as the positive input, which mimics the human behaviour of point selection. The parameter $K$ is set to 2. For the negative point, we randomly choose one from the background.

\subsection{Patch up Low-Level Semantic Information}

\begin{figure}[!tbp]
  \centering
  \includegraphics[width=0.7\linewidth]{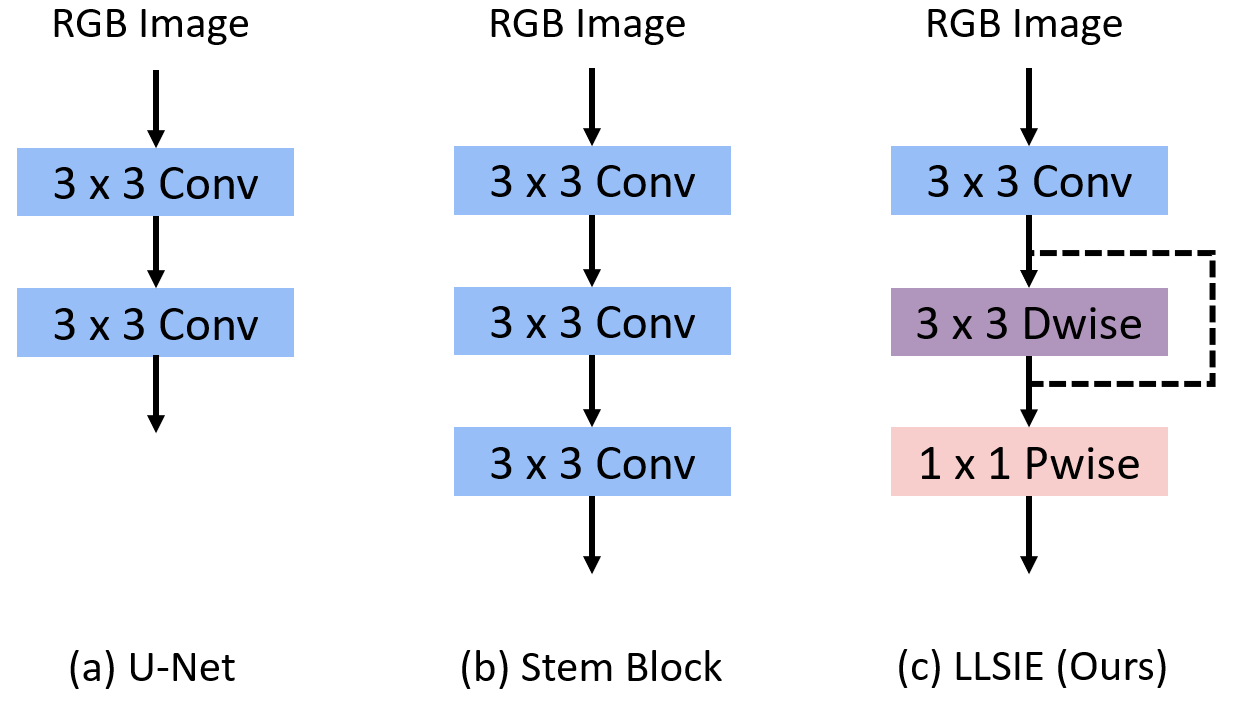}
  \caption{Comparison of our design and other common structures for low-level feature extraction from input images.}
  \label{fig:f2}
\end{figure}

For the image encoder, SAM adopts a large-sized MAE \cite{he2022masked} pre-trained ViT to extract features from input images. Although point prompt-based SAM has performed amazing zero-shot ability in natural image segmentation, medical image segmentation still can be a challenging task due to its complexity of interpretation. To retrain or fine-tune SAM on Nuclei datasets, 256 A100 GPUS are required in the training phase \cite{kirillov2023segment}. It can be a non-trivial burden for many researchers. Therefore, we replace SAM image encoder with a Tiny-ViT \cite{wu2022tinyvit}, which drops the number of parameters by about 99.1\%. However, such an operation somehow diminishes the feature extraction ability of the image encoder. Inspired by previous U-shape segmentation networks, we construct a low-level semantic information extractor (LLSIE) to patch up shallow features. The U-Net module \cite{ronneberger2015u} in Fig. \ref{fig:f2} (a) has been widely used in various architectures \cite{jha2020doubleu,huang2020unet}. The stem block \cite{chen2019mmdetection} in Fig. \ref{fig:f2} (b) is usually developed to obtain the larger receptive field. The structure of LLSIE is shown in Fig. \ref{fig:f2} (c). In order to decrease the 
computational cost and parameters of the network, our proposed module uses depthwise separable convolution \cite{howard2017mobilenets}, composed of $3\times3$ depthwise convolution followed by $1\times1$ pointwise convolution. The former is used to merge spatial information. The latter integrates channel information. A $3\times3$ general convolution is inserted into the head of LLSIE to preserve the receptive field and increase the feature dimension as depthwise separable convolution cannot perform well on low-channel feature maps \cite{sandler2018mobilenetv2}. In the end, we involve a residual connection to mitigate the potential effect of the gradient vanish.

\subsection{SPPNet Architecture}

\begin{figure}[!thbp]
  \centering
  \includegraphics[width=1\linewidth]{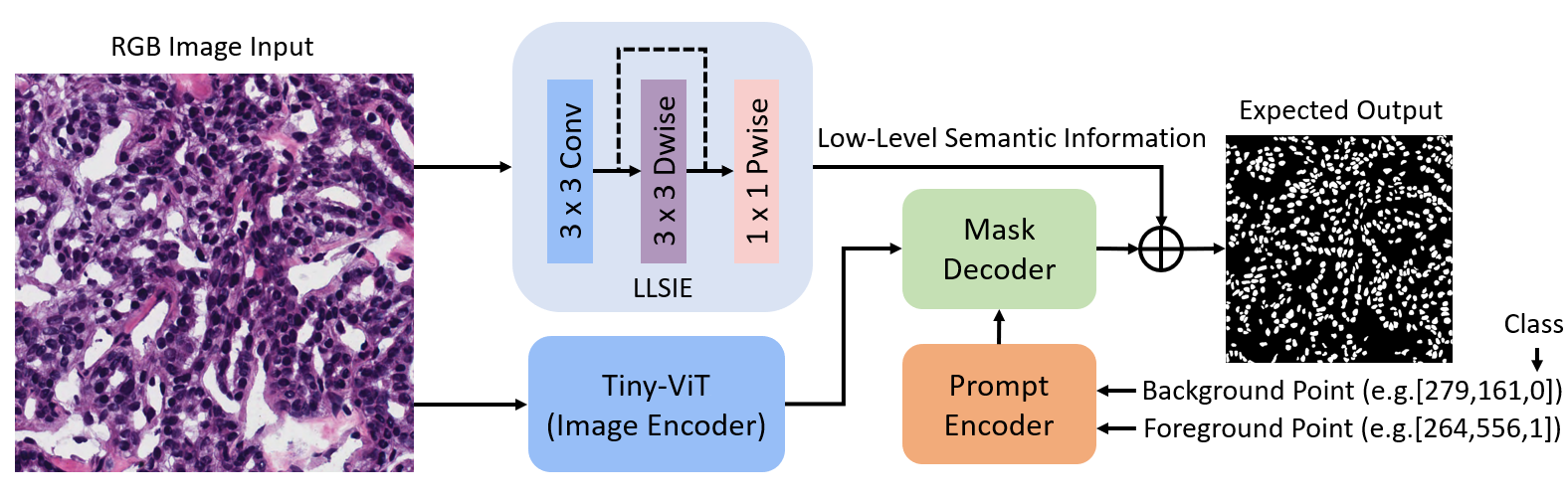}
  \caption{Overview of our single-point prompt network with efficient vison transformer.}
  \label{fig:f3}
\end{figure}

For nuclei image segmentation, we establish the SPPNet based on the proposed point sampling strategy and LLSIE block. Fig. \ref{fig:f3} displays an overview of our SSPNet.
The input RGB microscopic images are first fed to the Tiny-ViT for feature extraction. A single set of point prompts, including one cell point and one background point, will be generated using center neighborhood point sampling, and then used to collect tokens by prompt encoder. The mask decoder leverages multilayer perceptrons and a cross-attention mechanism to upgrade the image embedding as well as prompt, and raise the width and height of feature maps to $256 \times 256$. In addition, the LLSIE block is in parallel with the image encoder for capturing low-level semantic information from inputs. This block is followed by a $2\times2$ max pooling with stride 2 for aligning with the shape of the feature map produced by the mask decoder. A connection operation fuses both two features and one $1\times1$ convolution declines the channel to the same number of classes as prediction. Finally, we reuse the post-processing method of SAM to recover the image size and generate masks.

\section{Experiments and Results}

\subsection{Implementation Details}
All experiments are conducted with PyTorch 1.10.0 framework on a single NVIDI-A RTX2080Ti Tensor Core GPU, 4-core CPU, and 28GB RAM. A standard dice loss function and an Adam optimizer with a learning rate of 5e-4 are used to train all models. We set batch sizes and epochs to 4 and 200 respectively. The early stopping strategy is implemented to avoid overfitting. During training our SSPNet, we input original microscopic images to the image encoder and resized images ($256\times256$) to the LLSIE block. These resized images are also used to train other SOTA methods for comparison. Meanwhile, Tiny-ViT loads a pre-trained checkpoint based on knowledge distillation \cite{zhang2023faster}. The prompt encoder and mask decoder of SPPNet reuse the pre-trained weight of SAM \cite{kirillov2023segment}. We randomly select 80\% samples as a training set, 10\% as a validation set, and 10\% as a test set. Moreover, data augmentation methods have been widely used to extend image diversity and improve model robustness. In the experiment, horizontal flip, rotation, and cutout with a probability of 0.25 are randomly applied to the training set.

\subsection{Evaluation metrics.}
In order to evaluate and quantify mode performance. Mean intersection over union (mIoU) and dice coefficient (DSC), the standard medical image segmentation metrics, are computed on the test set. Both show the similarity between prediction and ground truth using different calculation methods, where mIoU tends to penalise single instances of bad classification. Additionally, we report parameters (Params), floating point operations (FLOPs), and frames per second (FPS) of models to reveal their sizes, computational complexity, and inference speed.

\subsection{Comparison on MoNuSeg-2018 dataset.}
Microscopic images usually include a large number of cells. Manual labeling is expensive and time-consuming. Therefore, we select a small dataset to evaluate all models, which is more feasible in clinical applications. MoNuSeg-2018 is an open-access dataset for MICCAI 2018 multi-organ nuclei segmentation challenge. It only contains 51 fully-labeled pathological images. We trained SPPNet and four SOTA models. The quantitative result of the test set is provided in Table \ref{tab1}.
\begin{table}[!htbp]
  \centering
  \setlength\tabcolsep{7pt}
  \caption{Performance comparison between our method (SSPNet) and other SOTA models on MoNuSeg-2018 dataset.}
  {\scalebox{0.93}{
  \begin{tabular}{llllll}
  \toprule
  Method & mIoU(\%) & DSC(\%) & Params(M) & FLOPs & FPS\\
  \midrule
  U-Net \cite{ronneberger2015u} & 61.64±8.28 & 75.92±6.79 & 13.40 & 23.83 & 93.22\\
  Unet++ \cite{zhou2019unet++} & 62.28±6.54 & 76.05±5.06 & 9.16 & 26.73 & 69.88\\
  Attention U-Net \cite{schlemper2019attention} & 63.85±7.23 & 77.68±5.82 & 34.88 & 51.02 & 55.19 \\
  TransUNet \cite{chen2021transunet} & 57.87±6.41 & 73.09±5.51 & 61.82 & 32.63 & 66.67\\
  Swin-UNet \cite{cao2022swin} & 58.21±6.50 & 73.36±5.50 & 27.15 & 5.91 & 43.12\\
  SPPNet & \textbf{66.43±4.32} & \textbf{79.77±3.11} & 9.79 & 39.90 & 22.61 \\
  \bottomrule
  \end{tabular}}}
  \label{tab1}
\end{table}
We can observe that SPPNet achieves a DSC of 79.77\% and a mIoU of 66.43\%, which outperforms Attention U-Net by 2.09\% in terms of DSC and 2.58\% in mIoU. Particularly, our proposed model demonstrates a significant enhancement over the two previous transformer-based architectures, where the mIoU of SPPNet is 8.56\% and 8.22\% higher than TransUNet and Swin-UNet, and the DSC of SPPNet is 6.68\% and 6.41\% higher than these two models respectively. Also, SPPNet costs considerably fewer parameters. Consequently, our proposed model reveals the highest score in standard evaluation metrics of medical image segmentation.

\subsection{Ablation Study}

In this section, a detailed ablation study is conducted on SPPNet. LLSIE block is an essential module to capture low-level semantic information from input images. We first compare LLSIE with other prevalent blocks, which is shown in Table \ref{tab2}. 
\begin{table}[!thbp]
  \centering
  \setlength\tabcolsep{7pt}
  \caption{Performance comparison between our LLSIE block and other general methods for low-level semantic information extraction on MoNuSeg-2018 dataset.}
  {\scalebox{0.7}{
  \begin{tabular}{llllll}
  \toprule
  Block & mIoU(\%) & DSC(\%) & Params(M) & FLOPs & FPS\\
  \midrule
  U-Net \cite{ronneberger2015u}& 64.95±7.94 & 77.70±6.37 & 9.80 & 40.41 & 22.12 \\
  Stem  & 65.13±6.69 & 78.93±5.19 & 9.81 & 41.02 & 21.55\\
  LLSIE & \textbf{66.43±4.32} & \textbf{79.77±3.11} & 9.79 & 39.90 & 22.61 \\
  \bottomrule
  \end{tabular}}}
  \label{tab2}
\end{table}
\begin{figure}[!thbp]
  \centering
  \includegraphics[width=0.7\linewidth]{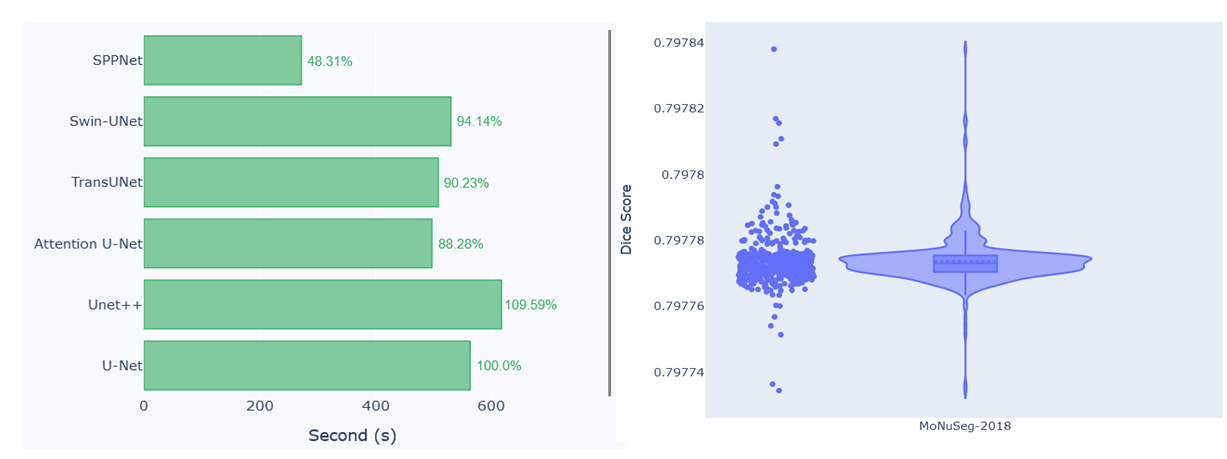}
  \caption{Training time cost to achieve SOTA performance with the U-Net baseline (left side) and presentation of model stability with violin plot (right side).}
  \label{fig:f4}
\end{figure}
It can be demonstrated that the LLSIE block not only performs better than U-Net and Stem blocks in metrics of mIoU and DSC, but also has faster inference speed and costs a lower number of parameters as well as computational resources. Furthermore, we evaluate the efficiency of Tiny-ViT, center neighborhood point sampling strategy, and LLSIE block, which is presented in Table \ref{tab3}. We make a comparison to SAM with fine-tuned prompt encoder and mask decoder. 
\begin{table}[!tbp]
  \centering
  \setlength\tabcolsep{7pt}
  \caption{Detailed ablation study of our single-point prompt network architecture on dataset. IE: image encoder, PE: prompt encoder, MD: mask decoder, CNPS: center neighborhood point sampling}
  {\scalebox{0.70}{
  \begin{tabular}{ccc|ccc|lllll}
  \toprule
  \multicolumn{3}{c}{SAM} & \multicolumn{3}{c}{SPPNet} & & & & &\\
  IE & PE & MD & IE & CNPS & LLSIE & mIoU(\%) & DSC(\%) & Params(M) & FLOPs & FPS\\
  \midrule
  \checkmark & \checkmark & \checkmark &  &  &  & 60.18±8.15 & 74.76±7.00 & 635.93 & 2736.63 & 1.39\\
   & \checkmark & \checkmark & \checkmark &  &  & 62.33±4.05 & 76.22±2.98 & 9.78 & 39.73 & 23.78\\ 
   &  \checkmark & \checkmark & \checkmark & \checkmark &  & 63.77±5.77 & 77.73±4.26 & 9.78 & 39.73 & 23.64\\
   &  \checkmark & \checkmark & \checkmark & \checkmark & \checkmark & \textbf{66.43±4.32} & \textbf{79.77±3.11} & 9.79 & 39.90 & 22.61 \\
  \bottomrule
  \end{tabular}}}
  \label{tab3}
\end{table}
From Table \ref{tab3}, we can first conclude that it is necessary to retrain the image encoder when applied to the nuclei segmentation task as retrained Tiny-ViT outperforms pre-trained MAE ViT. Secondly, our center neighborhood point sampling strategy can improve model performance without any extra parameters and computational costs. Thirdly, incorporating low-level semantic information can help the model predict a more precise segmentation mask. Overall, compared to SAM, SPPNet shows better performance, considerably fewer parameters, lower computational costs, and faster inference.

\begin{figure*}[!thbp]
  \centering
  \includegraphics[width=0.9\linewidth]{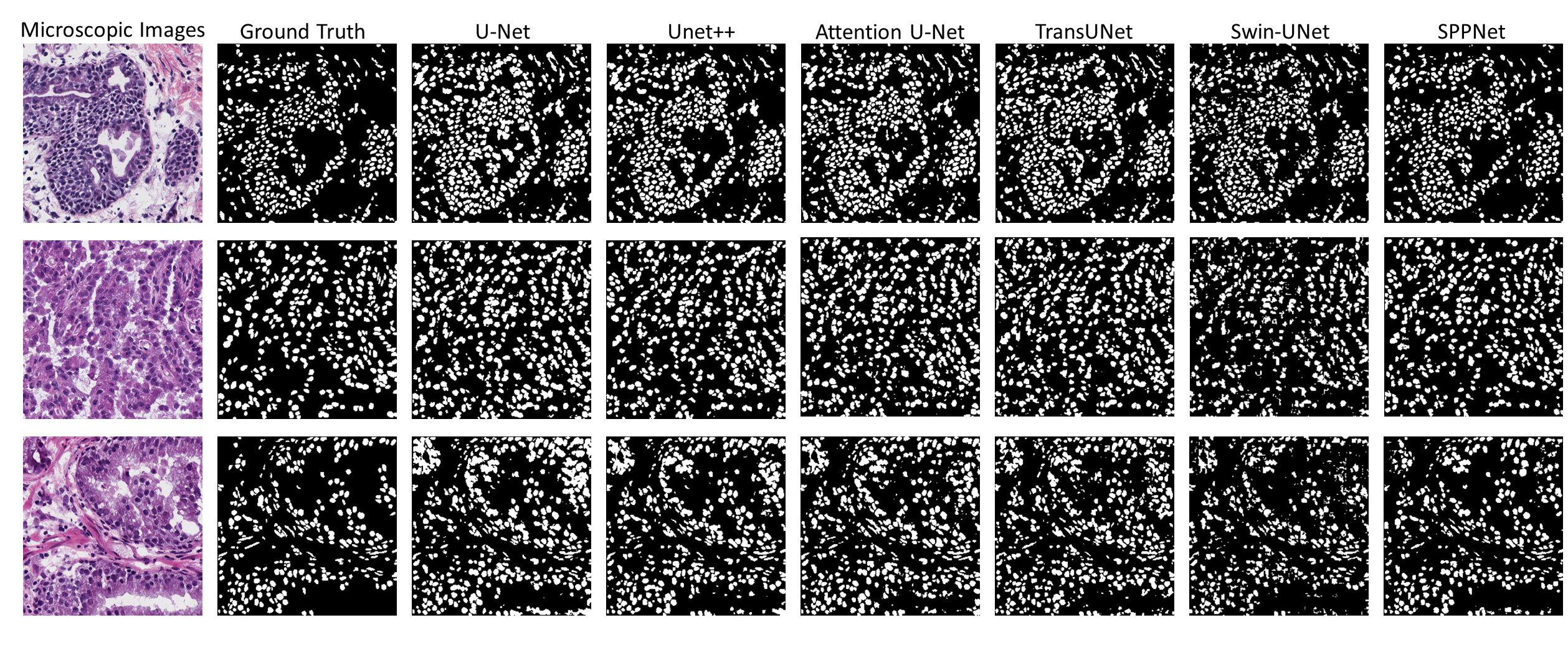}
  \caption{Qualitative comparison results between SPPNet and other SOTA models on the test set of MoNuSeg-2018.}
  \label{fig:f5}
\end{figure*}

\subsection{Model Stability}

When applying the center neighborhood point sampling strategy to determine a positive point, the target cell and neighbors are randomly selected by our program. To evaluate our model stability, we use the same trained model to implement the test set with 500 iterations. The result is provided in Fig. \ref{fig:f4} (right side) with a violin plot. It can be demonstrated that the dice score of the proposed SPPNet is clustered at 79.776\% to 79.778\%. The maximum performance gap is only about 0.01\%. As a result, we can argue that the random point selection based on our sampling strategy does not have a significant impact on the model performance.

\subsection{Dicussion}

Nuclei image segmentation is an important step for pathological analysis for patients. From the above experimental results, our approach shows an outstanding improvement compared to SAM in training resource costs and better performance than existing SOTA models of medical image segmentation. Also, SPPNet can take about 50\% less time than others in the training phase, which is shown in Fig. \ref{fig:f4} (left side).  To further demonstrate the efficiency of SPPNet in clinical applications, some segmentation masks using all models on the test set are visualised in Fig. \ref{fig:f5}. Compared to other architectures, SPPNet is able to provide more accurate masks and fewer false positive cases. In addition, SAM displays a great zero-shot ability in natural image segmentation as it is trained on an extremely sufficient dataset. Therefore, we will also explore the zero-shot performance of SPPNet for nuclei image segmentation in the future. Achieving a such foundation model can considerably alleviate the paucity of pathologists.

\section{Conclusion}
In this paper, we propose a single-point prompt network for nuclei image segmentation. The LLSIE block is used to patch up lower-level semantic information in the mask decoder. A center neighborhood point sampling strategy makes inference more feasible in clinical applications. The experimental results demonstrate that our SPPNet can be considered a new benchmark for cell segmentation. 

%
%
%
\bibliographystyle{splncs04}
\bibliography{refs}

\begin{thebibliography}{10}
\providecommand{\url}[1]{\texttt{#1}}
\providecommand{\urlprefix}{URL }
\providecommand{\doi}[1]{https://doi.org/#1}

\bibitem{cao2022swin}
Cao, H., Wang, Y., Chen, J., Jiang, D., Zhang, X., Tian, Q., Wang, M.:
  Swin-unet: Unet-like pure transformer for medical image segmentation. In:
  European conference on computer vision. pp. 205--218. Springer (2022)

\bibitem{chen2021transunet}
Chen, J., Lu, Y., Yu, Q., Luo, X., Adeli, E., Wang, Y., Lu, L., Yuille, A.L.,
  Zhou, Y.: Transunet: Transformers make strong encoders for medical image
  segmentation. arXiv preprint arXiv:2102.04306  (2021)

\bibitem{chen2019mmdetection}
Chen, K., Wang, J., Pang, J., Cao, Y., Xiong, Y., Li, X., Sun, S., Feng, W.,
  Liu, Z., Xu, J., et~al.: Mmdetection: Open mmlab detection toolbox and
  benchmark. arXiv preprint arXiv:1906.07155  (2019)

\bibitem{dosovitskiy2020image}
Dosovitskiy, A., Beyer, L., Kolesnikov, A., Weissenborn, D., Zhai, X.,
  Unterthiner, T., Dehghani, M., Minderer, M., Heigold, G., Gelly, S., et~al.:
  An image is worth 16x16 words: Transformers for image recognition at scale.
  arXiv preprint arXiv:2010.11929  (2020)

\bibitem{he2022masked}
He, K., Chen, X., Xie, S., Li, Y., Doll{\'a}r, P., Girshick, R.: Masked
  autoencoders are scalable vision learners. In: Proceedings of the IEEE/CVF
  conference on computer vision and pattern recognition. pp. 16000--16009
  (2022)

\bibitem{he2023accuracy}
He, S., Bao, R., Li, J., Grant, P.E., Ou, Y.: Accuracy of segment-anything
  model (sam) in medical image segmentation tasks. arXiv preprint
  arXiv:2304.09324  (2023)

\bibitem{howard2017mobilenets}
Howard, A.G., Zhu, M., Chen, B., Kalenichenko, D., Wang, W., Weyand, T.,
  Andreetto, M., Adam, H.: Mobilenets: Efficient convolutional neural networks
  for mobile vision applications. arXiv preprint arXiv:1704.04861  (2017)

\bibitem{huang2020unet}
Huang, H., Lin, L., Tong, R., Hu, H., Zhang, Q., Iwamoto, Y., Han, X., Chen,
  Y.W., Wu, J.: Unet 3+: A full-scale connected unet for medical image
  segmentation. In: ICASSP 2020-2020 IEEE international conference on
  acoustics, speech and signal processing (ICASSP). pp. 1055--1059. IEEE (2020)

\bibitem{jha2020doubleu}
Jha, D., Riegler, M.A., Johansen, D., Halvorsen, P., Johansen, H.D.:
  Doubleu-net: A deep convolutional neural network for medical image
  segmentation. In: 2020 IEEE 33rd International symposium on computer-based
  medical systems (CBMS). pp. 558--564. IEEE (2020)

\bibitem{kirillov2023segment}
Kirillov, A., Mintun, E., Ravi, N., Mao, H., Rolland, C., Gustafson, L., Xiao,
  T., Whitehead, S., Berg, A.C., Lo, W.Y., et~al.: Segment anything. arXiv
  preprint arXiv:2304.02643  (2023)

\bibitem{kumar2017dataset}
Kumar, N., Verma, R., Sharma, S., Bhargava, S., Vahadane, A., Sethi, A.: A
  dataset and a technique for generalized nuclear segmentation for
  computational pathology. IEEE transactions on medical imaging
  \textbf{36}(7),  1550--1560 (2017)

\bibitem{ronneberger2015u}
Ronneberger, O., Fischer, P., Brox, T.: U-net: Convolutional networks for
  biomedical image segmentation. In: International Conference on Medical image
  computing and computer-assisted intervention. pp. 234--241. Springer (2015)

\bibitem{sandler2018mobilenetv2}
Sandler, M., Howard, A., Zhu, M., Zhmoginov, A., Chen, L.C.: Mobilenetv2:
  Inverted residuals and linear bottlenecks. In: Proceedings of the IEEE
  conference on computer vision and pattern recognition. pp. 4510--4520 (2018)

\bibitem{schlemper2019attention}
Schlemper, J., Oktay, O., Schaap, M., Heinrich, M., Kainz, B., Glocker, B.,
  Rueckert, D.: Attention gated networks: Learning to leverage salient regions
  in medical images. Medical image analysis  \textbf{53},  197--207 (2019)

\bibitem{wang2022medical}
Wang, R., Lei, T., Cui, R., Zhang, B., Meng, H., Nandi, A.K.: Medical image
  segmentation using deep learning: A survey. IET Image Processing
  \textbf{16}(5),  1243--1267 (2022)

\bibitem{wu2022tinyvit}
Wu, K., Zhang, J., Peng, H., Liu, M., Xiao, B., Fu, J., Yuan, L.: Tinyvit: Fast
  pretraining distillation for small vision transformers. In: European
  Conference on Computer Vision. pp. 68--85. Springer (2022)

\bibitem{yan2022after}
Yan, X., Tang, H., Sun, S., Ma, H., Kong, D., Xie, X.: After-unet: Axial fusion
  transformer unet for medical image segmentation. In: Proceedings of the
  IEEE/CVF winter conference on applications of computer vision. pp. 3971--3981
  (2022)

\bibitem{zhang2023faster}
Zhang, C., Han, D., Qiao, Y., Kim, J.U., Bae, S.H., Lee, S., Hong, C.S.: Faster
  segment anything: Towards lightweight sam for mobile applications. arXiv
  preprint arXiv:2306.14289  (2023)

\bibitem{zhou2019unet++}
Zhou, Z., Siddiquee, M.M.R., Tajbakhsh, N., Liang, J.: Unet++: Redesigning skip
  connections to exploit multiscale features in image segmentation. IEEE
  transactions on medical imaging  \textbf{39}(6),  1856--1867 (2019)

\end{thebibliography}
\end{document}